\documentclass[preprint,12pt]{elsarticle}

\journal{Nuclear Instruments and Methods in Physics Research A}

\begin{document}

\begin{frontmatter}

\title{First FBK Production of 50$\mu$m Ultra-Fast Silicon Detectors}

\author[unito,infn]{V. Sola\corref{corrauth}}
\cortext[corrauth]{Corresponding author}
\ead{valentina.sola@cern.ch}

\author[upo,infn]{R. Arcidiacono}
\author[fbk,tifpa]{M. Boscardin}
\author[infn]{N. Cartiglia}
\author[unitn,tifpa]{G.-F. Dalla Betta}
\author[fbk,tifpa]{F. Ficorella}
\author[unito,infn]{M. Ferrero}
\author[infn]{M. Mandurrino}
\author[unitn,tifpa]{L. Pancheri}
\author[fbk,tifpa]{G. Paternoster}
\author[infn]{A. Staiano}

\address[unito]{Universit\`a degli Studi di Torino, via P. Giuria 1, 10125, Torino, Italy}
\address[infn]{INFN, Sezione di Torino, via P. Giuria 1, 10125, Torino, Italy}
\address[upo]{Universit\'a del Piemonte Orientale, largo Donegani 2/3, 28100, Novara, Italy}
\address[fbk]{Fondazione Bruno Kessler, via Sommarive 18, 38123, Povo (TN), Italy}
\address[tifpa]{TIFPA-INFN, via Sommarive 18, 38123, Povo (TN), Italy}
\address[unitn]{Universit\`a degli Studi di Trento, via Sommarive 9, 38123, Povo (TN), Italy}

\begin{abstract}
  Fondazione Bruno Kessler (FBK, Trento, Italy) has recently delivered its first 50 $\mu$m thick production of
  Ultra-Fast Silicon Detectors (UFSD), based on the Low-Gain Avalanche Diode design.~These sensors use high
  resistivity Si-on-Si substrates, and have a variety of gain layer doping profiles and designs based on Boron,
  Gallium, Carbonated Boron and Carbonated Gallium to obtain a controlled multiplication mechanism.~Such variety of
  gain layers will allow identifying the most radiation hard technology to be employed in the production of UFSD,
  to extend their radiation resistance beyond the current limit of $\phi \sim$ 10$^{15}$ n$_{eq}$/cm$^2$.~In this paper,
  we present the characterisation, the timing performance, and the results on radiation damage tolerance of this
  new FBK production.
\end{abstract}

\begin{keyword}
  Silicon \sep Fast detector \sep Low gain \sep Charge multiplication \sep LGAD
\end{keyword}

\end{frontmatter}



\section{Introduction}

Ultra-Fast Silicon Detectors (UFSD) \cite{UFSD1,UFSD2,UFSDref}
are innovative silicon sensors optimised for timing measurements based on the Low-Gain Avalanche Diode (LGAD) \cite{LGAD1} design.
LGAD merges the best characteristics of traditional silicon sensors with the main feature of Avalanche Photodiodes (APD), using
n-in-p silicon diodes with a low and controlled internal multiplication mechanism. To obtain charge multiplication, an electric field of
the order of E $\sim$ 300 kV/cm is required, under this condition the electrons (and to less extent the holes) acquire enough
kinetic energy that are able to generate additional e/h pairs. The required field value is obtained by implanting an appropriate
acceptor density (N$_A \sim$10$^{16}$/cm$^3$) under the n$^{++}$ cathode.~The thickness of the acceptor implant is of the order of 1 $\mu$m.

UFSD recently obtained a time resolution of $\sigma_t \sim$ 30 ps in beam tests \cite{UFSD3} and
are now being considered in the upgrade of the CMS and ATLAS experiments as timing detectors~\cite{CMS-MTD,ATL-HGT}.

After the successful production of the first batch of 300 $\mu$m thick UFSD in 2016,
Fondazione Bruno Kessler (FBK) delivered its first 50 $\mu$m thick UFSD.~The 50 $\mu$m production
includes a variety of doping profiles and strategies of gain layers to identify the most radiation hard technology
for UFSD, to enhance their radiation tolerance beyond the current limit of $\phi \sim$ 10$^{15}$ n$_{eq}$/cm$^2$~\cite{HPKirr}.

\section{Production of 50 $\mathrm{\mu}$m thick UFSD at FBK}

The first production of UFSD at FBK was completed in 2016 on a 300~$\mu$m substrate (referred to as FBK UFSD1), and included several type of structures,
having both AC and DC-coupled devices~\cite{FBK_UFSD1a,FBK_UFSD1b}.~The goals of the production were the good control of the low gain mechanism
and the good correspondence between measurements and simulation of the gain as a function of the gain layer doping concentration.~The agreement
between data and simulation proved the success of this first production.

In 2017 FBK applied the know-how gained through the 300 $\mu$m production to thinner substrates, and manufactured UFSD on 60 $\mu$m thick Si-on-Si
substrates 6 inch wafers (namely FBK UFSD2).~After thermally bonding the high resistivity ($>$ 3000 Ohm$\cdot$cm) p-type Float-Zone
wafers to the 500 $\mu$m thick supports, the device active thickness reduces by $\sim$ 5 $\mu$m.~The
main goals of the 2017 production were to establish a reliable design for UFSD on thin substrates and to
test solutions to enhance the radiation tolerance of UFSD devices.

\begin{table}[b!]
\begin{center}
\begin{tabular}{ | c | c | c | c | c | } 
  \hline
  Wafer $\#$ & Dopant & Dose & Carbon & Diffusion \\
  \hline
   1 & Boron   & 0.98 &      & Low  \\
   2 & Boron   & 1.00 &      & Low  \\
   3 & Boron   & 1.00 &      & High \\
   4 & Boron   & 1.00 & Low  & High \\
   5 & Boron   & 1.00 & High & High \\
   6 & Boron   & 1.02 & Low  & High \\
   7 & Boron   & 1.02 & High & High \\  
   8 & Boron   & 1.02 &      & High \\
   9 & Boron   & 1.02 &      & High \\
  10 & Boron   & 1.04 &      & High \\
  \hline
  11 & Gallium & 1.00 &      & Low  \\
  12 & Gallium & 1.00 &      & Low  \\
  13 & Gallium & 1.04 &      & Low  \\
  14 & Gallium & 1.04 &      & Low  \\
  15 & Gallium & 1.04 & Low  & Low  \\
  16 & Gallium & 1.04 & High & Low  \\
  18 & Gallium & 1.08 &      & Low  \\
  19 & Gallium & 1.08 &      & Low  \\
  \hline
\end{tabular}
\caption{Summary of gain layer implant strategy}
\label{tab:split}
\end{center}
\end{table}

Recent radiation damage studies on UFSD from different manufacturers~\cite{GK-Irr1} show that the doping of the gain layer becomes
progressively deactivated by irradiation. In particular, it has been shown that the boron atoms are still present in the gain layer
volume, but they switch from substitutional to interstitial, not contributing to the gain mechanism anymore. Within the RD50 Collaboration
\cite{RD50}, it has been proposed to replace the Boron of the gain layer with Gallium, as Gallium might have a lower probability to become
interstitial than Boron~\cite{Ga1,Ga2}. A further proposal has been to add Carbon atoms in the gain layer volume to reduce the disappearance
of gain, as Carbon might slow down the acceptor removal mechanism protecting the Boron, or Gallium, dopant~\cite{C-YS}. Moreover,
the idea that reducing the implant volume could reduce the cross section of gain layer inactivation was proposed.~All these solutions have
been tested in the 50 $\mu$m FBK production, following a complete simulation of Gallium and Carbon implantation and diffusion.

Therefore, in the latest 50 $\mu$m thick UFSD production at FBK we have simulated and manufactured devices with 5 different gain layer
configurations, summarised in Table~\ref{tab:split}:
(i) Boron Low Diffusion, (ii) Boron, (iii) Boron with Carbon, (iv) Gallium, and (v) Gallium with Carbon.~The
Boron implants have 4 doping doses, where 1.00 is the reference implant dose as used in FBK UFSD1 production~\cite{FBK_UFSD1b}, 
and other splits are obtained in steps of 2$\%$ from the reference dose.~Two different diffusion temperatures (Low and High)
have been used, to vary the implant volume of the gain layer.~The Gallium have 3 implant doses of the gain layer
(step of 4$\%$, starting from the reference dose).~Only one diffusion temperature
was used, same as Boron Low, given the higher diffusivity of Gallium.~The Carbon implants have 2 doses of concentration
(Low and High), and was implanted inside the gain layer volume
prior than Boron or Gallium.~No thermal annealineg was performed to Carbon before the acceptor implantation.

The bulk dopant concentration is $\sim$ 3$\cdot$10$^{12}$/cm$^3$.

\section{Characterisation of FBK UFSD2 Production}

Extensive quality tests and electrical characterisation on wafers have been made at FBK.~The studies
concentrated on single pad sensors with an active area of 1~$\times$~1~mm$^2$.

\subsection{Current versus Reverse Bias}
\label{sec:IV}

\begin{figure}[t!]
\centering
\includegraphics[width=\linewidth]{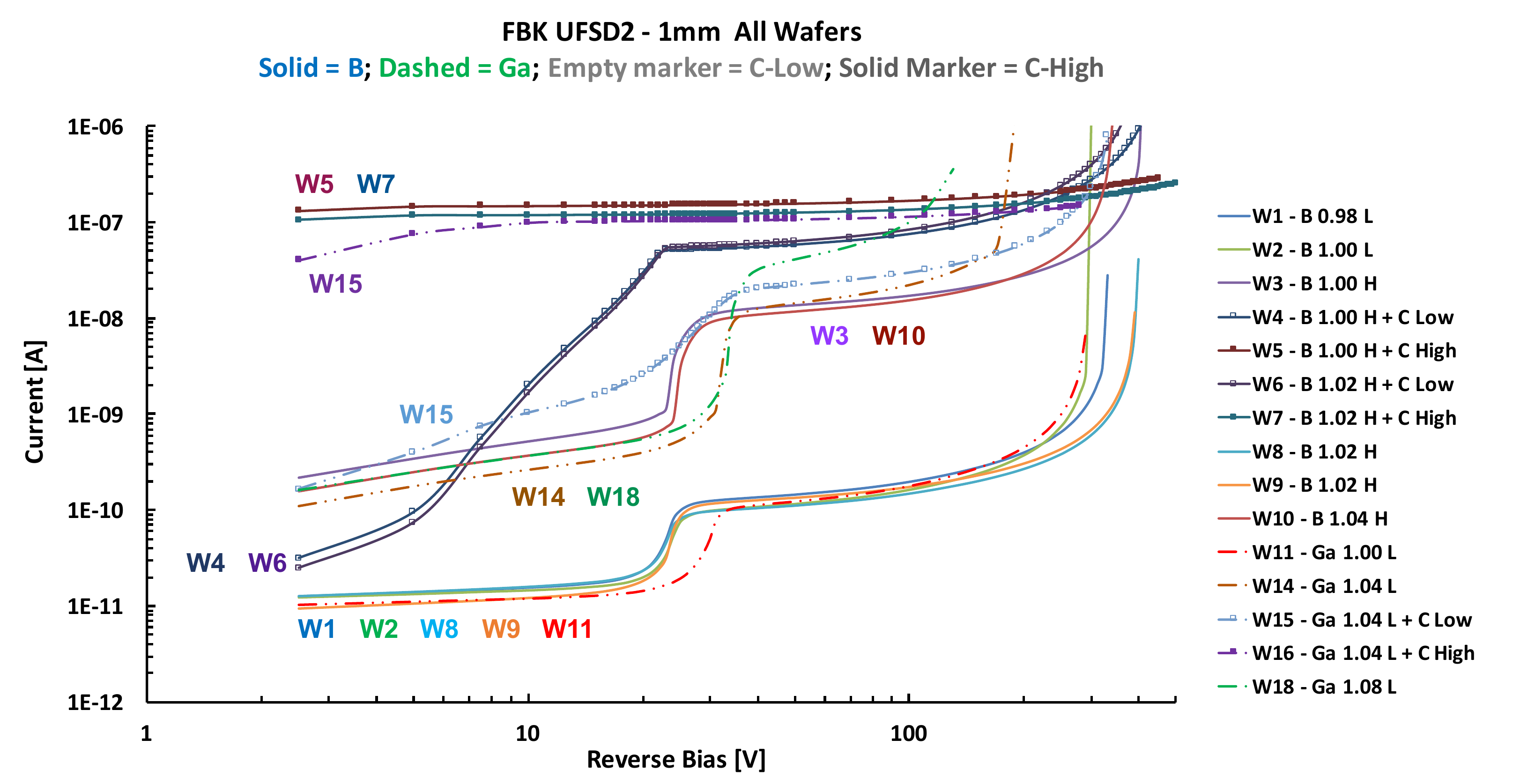}
\caption{Dark current as a function of the applied reverse bias for all the wafers produced in UFSD2. Solid lines indicate wafers with Boron,
  dashed lines wafers with Gallium; open squares indicate Low Carbon, solid squares High Carbon adjunction in the gain layer volume.}
  \label{fig:IVall}
\end{figure}

Figure~\ref{fig:IVall} shows the dark current as a function of the reverse bias measured at room temperature, on sensors from all the produced wafers.
Below 20 V the surface current is measured; between 20 and 30 V the current shows a rapid increase (ankle) due to the depletion of the gain layer 
and $\sim$ 5 V later (knee) flattens due to the depletion of the bulk; above 30 V sensors are fully depleted and the exponential growth of the
dark current is due to the relationship between gain and
the reverse bias~\cite{ImpIon}, as the charge multiplication follows:
\begin{equation}
  N(x) = N_0 \cdot e^{\alpha x} = N_0 \cdot G ~,
\label{eq:gain}
\end{equation}
where $x$ is the length travelled by the charge and G is the gain as function of $\alpha$, the impact ionisation rate,
which depends exponentially on the electric field E.
The two families of curves separated by a factor $\sim$ 100 are due to different production batches of the Si-on-Si wafers,
namely $good$ with low dark current and $leak$ with high dark current as listen on Table.~\ref{tab:batch},
and not to differences in the production process.

\begin{table}[b!]
\begin{center}
\begin{tabular}{ | c | c | } 
  \hline
  Batch & Wafer $\#$ \\
  \hline
  Good & W1, W2, W4, W6, W7, W8, W9, W11, W12, W13, W17, W19 \\
  \hline
  Leak & W3, W5, W10, W14, W15, W16, W18  \\
  \hline
\end{tabular}
\caption{List of wafers belonging to good or high leakage production batches.}
\label{tab:batch}
\end{center}
\end{table}

\begin{figure}[t!]
\centering
\includegraphics[width=0.8\linewidth]{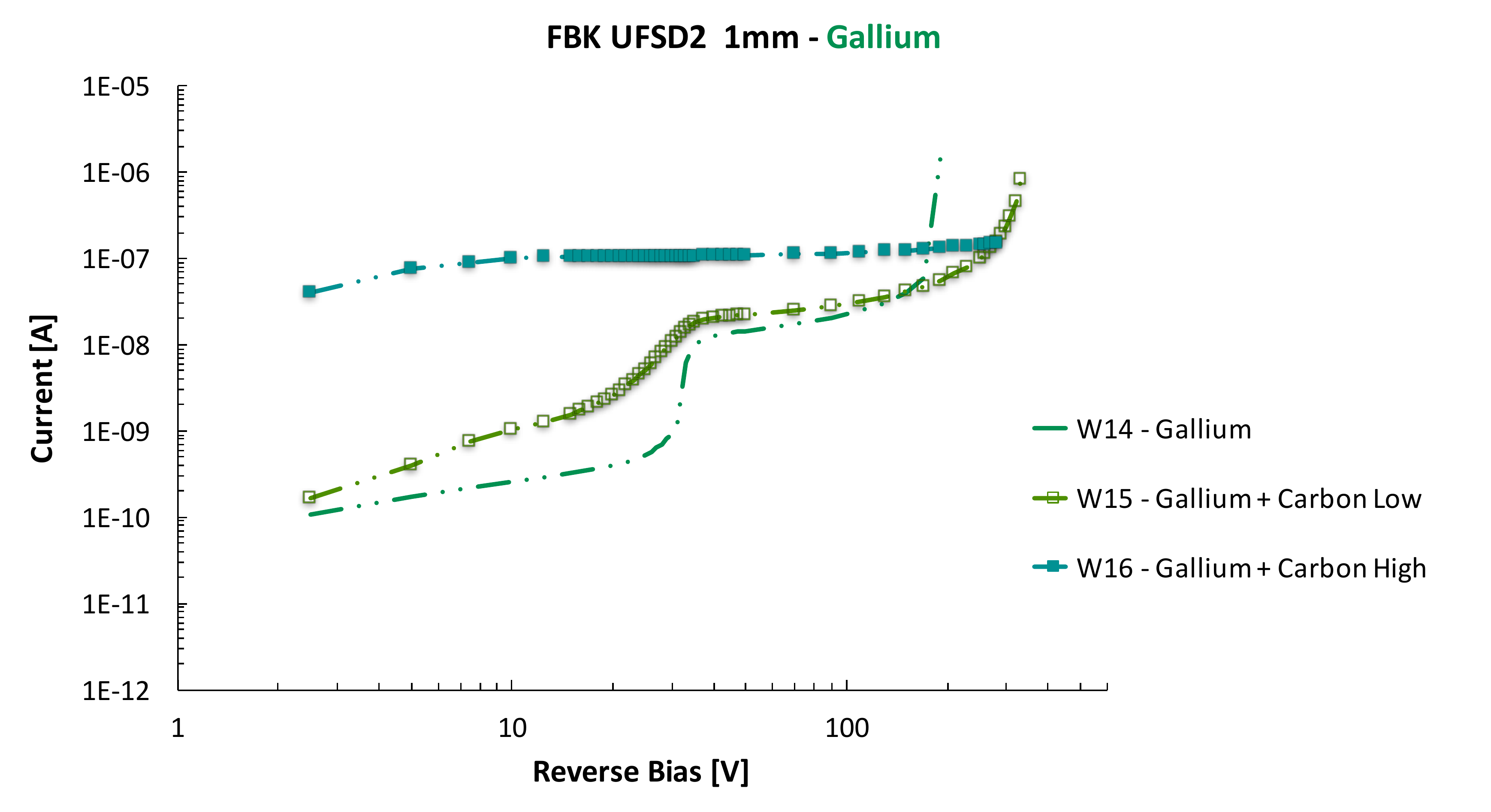}
\caption{Dark current as a function of the applied reverse bias for 3 different configurations of Gallium doped wafers:
  without Carbon (no markers), Low Carbon (open squares), High Carbon (solid squares).~Both No-Carbon and Low-Carbon
  sensors come from the high leakage batch of wafers.}
  \label{fig:IVga}
\end{figure}

More in detail, Fig.~\ref{fig:IVga} shows the effects of Carbon on wafers doped with Gallium.~An
implant of Low-Carbon dose does not affect the electrical
characteristics of the sensors, but slightly increases the dark current and shifts the depletion of the gain layer towards lower voltages.~On
the other hand, the High-Carbon dose sensor is dominated by the dark current, with an almost constant current.~Wafers with Boron gain implant
show very similar behaviour.

\begin{figure}[b!]
\centering
\includegraphics[width=0.8\linewidth]{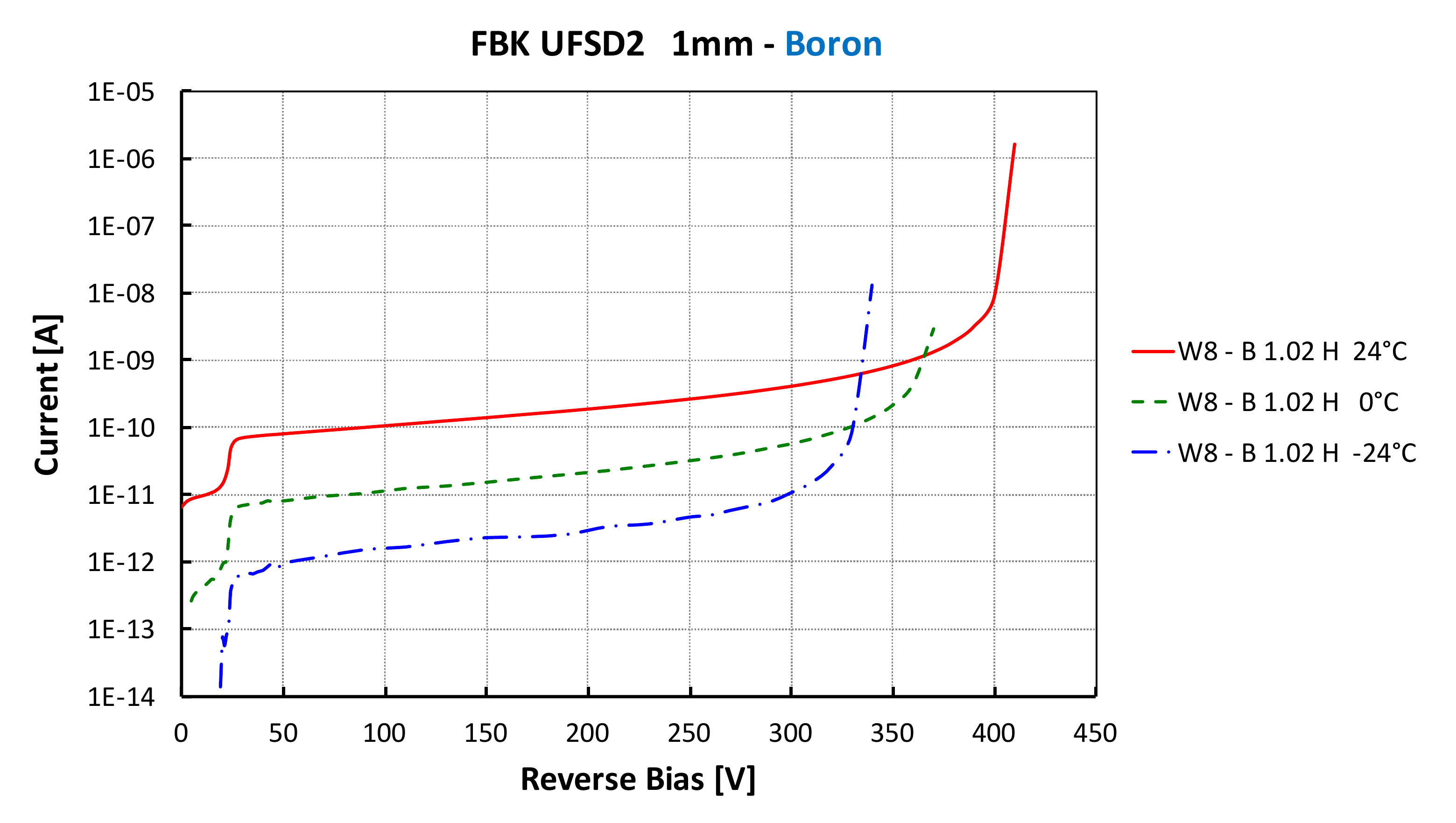}
\caption{Dark current as a function of the applied reverse bias of Wafer 8 measured at 3 different temperatures:
  $+24^{\circ}$C (plain line), $0^{\circ}$C (dashed line), $-24^{\circ}$C (dashed-dotted line).}
  \label{fig:IVw8t}
\end{figure}

Dark current measurements at different temperatures have been performed.~Decreasing the operating temperature of the sensors
reduces the overall dark current while increasing the internal gain of UFSD (for more details, see~\cite{TempRM}), as $G$ from Eq.~(\ref{eq:gain})
is influenced by the temperature by means of the carriers saturated velocities and the exponential temperature dependence of
the impact ionisation rate, $\alpha$.~This effect is responsible for a earlier sensor breakdown, as clearly visible in Fig.~\ref{fig:IVw8t} for Wafer 8.

\subsection{Capacitance versus Reverse Bias}
\label{sec:CV}

\begin{figure}[b!]
\centering
\includegraphics[width=\linewidth]{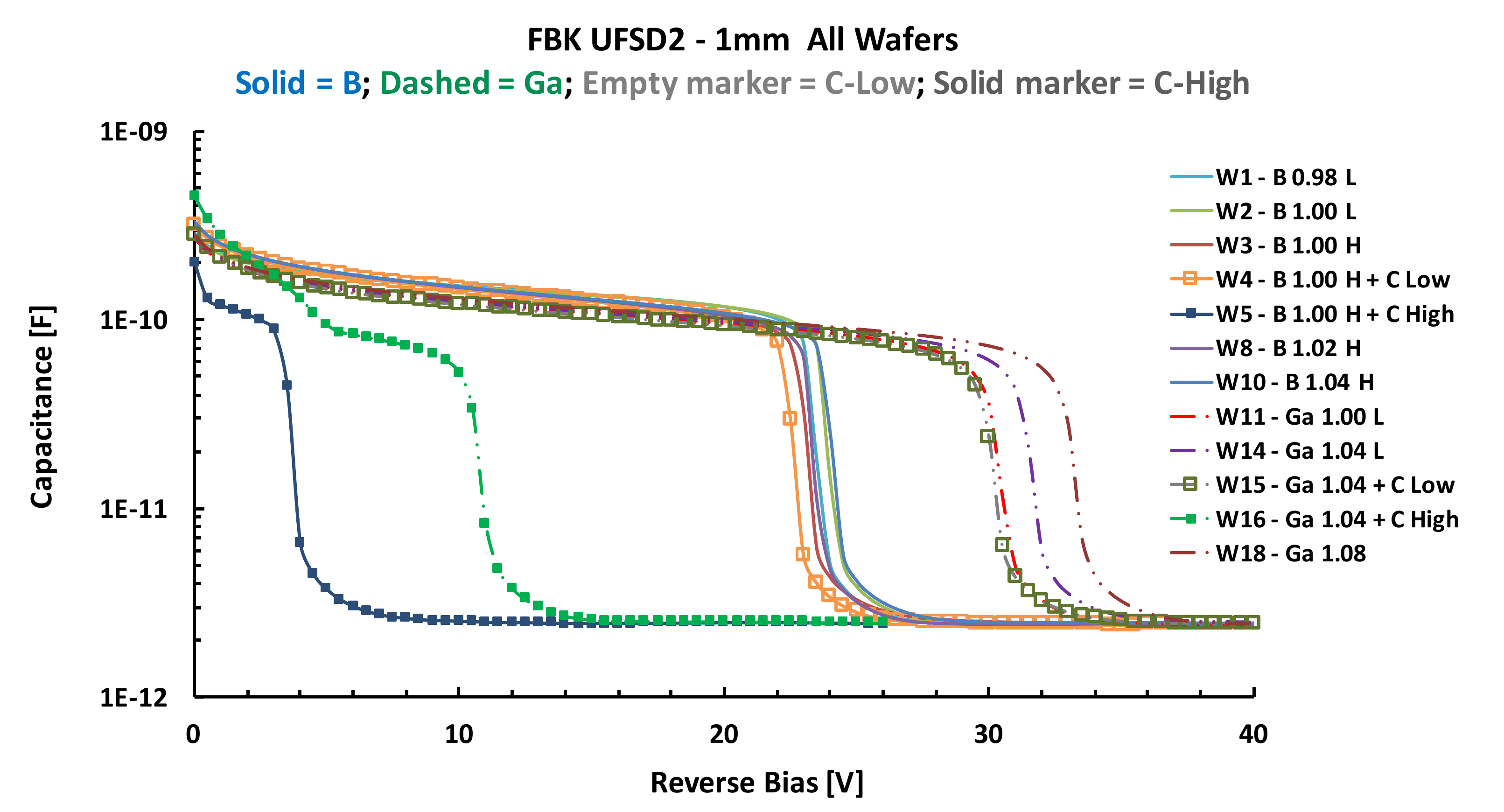}
\caption{Capacitance as a function of the applied reverse bias for sensors with different gain layer configurations.~Solid
  lines indicate wafers with Boron, dashed lines wafers with Gallium; open squares indicate Low Carbon, solid squares
  High Carbon adjunction in the gain layer volume.}
  \label{fig:CVall}
\end{figure}

Figure~\ref{fig:CVall} shows the capacitance as a function of the reverse bias measured at room temperature, for sensors with different gain layer
configurations. Similarly to what already observed in Fig.~\ref{fig:IVall}, C-V curves show a drop at a voltage corresponding to the
depletion of the gain layer, while the point at which the curves flatten indicates the full depletion of the sensor,
when also the high-resistive bulk is depleted.

As for standard silicon sensors, UFSD with a reverse bias can be considered as a parallel plate capacitor, where the distance between the capacitor
plates is the width of the depletion region, and the area of the plates is the active area of the sensors. The capacitance $C$, of the sensor is
related to the reverse bias $V$, and the acceptor active doping concentration $N_A$, as given by the relation:
\begin{equation}
  C \propto \sqrt{N_A/V}~.
\label{eq:cv}
\end{equation}
Therefore, assuming the same purity of the bulk for all wafers, the point at which the capacitance drops is directly related to the active doping
concentration of the gain layer.

From Fig.~\ref{fig:CVall}, it is possible to distinguish the family of Gallium doped sensors (dashed lines), with different doping concentration,
and the family of Boron doped sensors (plain lines), as listed in Table~\ref{tab:split}.~Moreover, the effect of Carbon implantation is visible:
for Low Carbonated sensors (open squares) a small reduction of the active doping concentration is noticeable, while for High Carbonated sensors
the active doping concentration of the gain layer is dramatically reduced.

\subsection{Acceptor Doping Concentration versus Depth}
\label{sec:DC}

\begin{figure}[b!]
\centering
\includegraphics[width=0.8\linewidth]{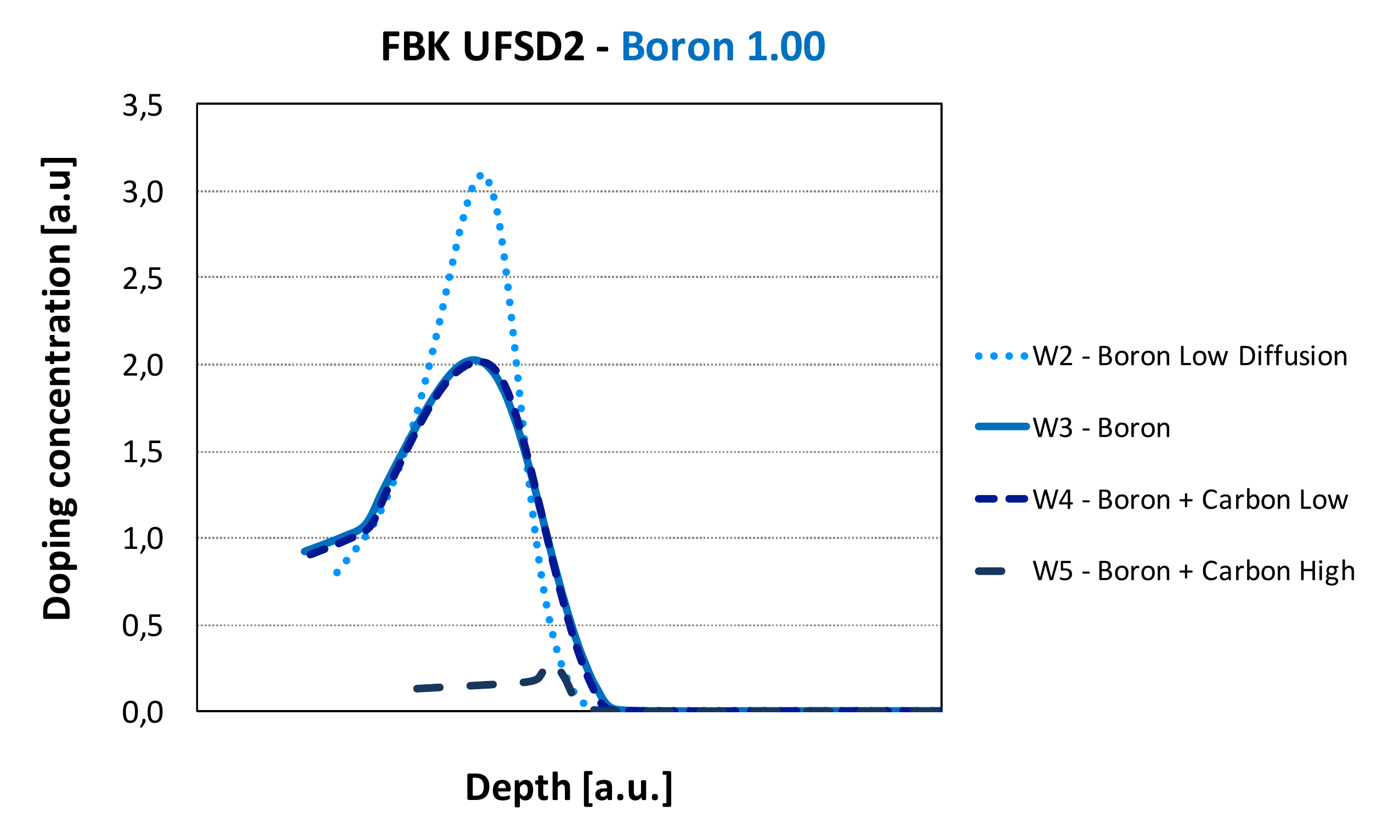}
\caption{Active dopant concentration as a function of the UFSD depth for Boron doped gain implants, comparing Boron only,
  Low Diffusion (dotted line) and standard Diffusion (plain line), Boron plus Low Carbon (dashed line) and Boron plus High Carbon
  (dotted-dashed line).}
  \label{fig:DPb}
\end{figure}

\begin{figure}[t!]
\centering
\includegraphics[width=0.8\linewidth]{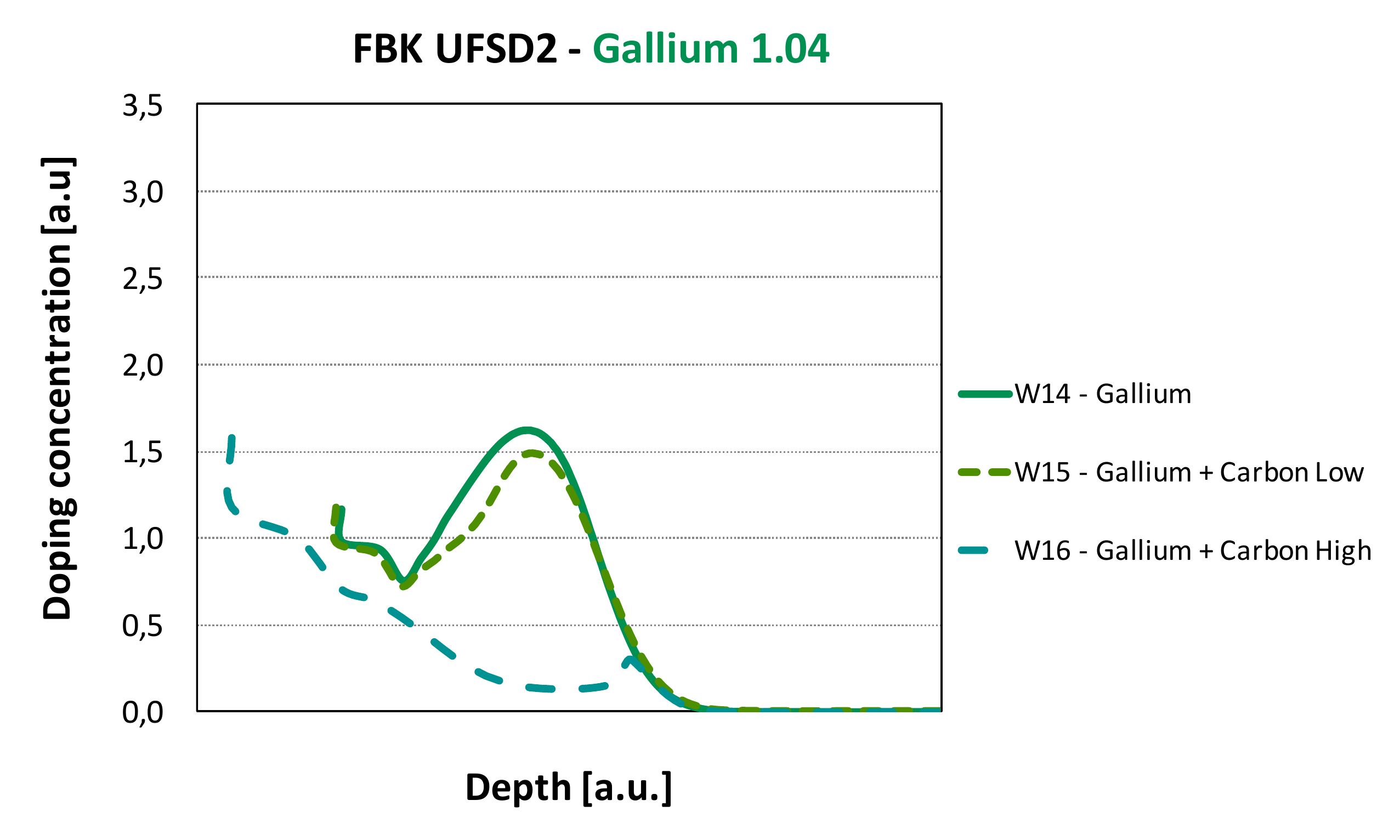}
\caption{Active dopant concentration as a function of the UFSD depth for Gallium doped gain implants, comparing Gallium only
  (plain line), Gallium plus Low Carbon (dashed line) and Gallium plus High Carbon (dotted-dashed line).}
  \label{fig:DPga}
\end{figure}

From the measurements of capacitance presented in Sect.~\ref{sec:CV} and exploiting the relation
\begin{equation}
  N(w) = \frac{2}{q \epsilon_{Si} A} \frac{1}{d(1/C^2)/dV}~~
  \mathrm{where}~~
  w = \frac{\epsilon_{Si} A^2}{C(V)}
\label{eq:doping}
\end{equation}
it is possible to extract the active doping density, $N$, as a function of the depletion depth, $w$,
being $A$ the sensor area and $\epsilon_{Si}$ the Silicon permittivity.

Figures~\ref{fig:DPb}~and~\ref{fig:DPga} show the doping densities as a function of the sensor depth for Boron and Gallium based gain implants,
respectively (plots are on scale). The origin of the x-axes corresponds to the p-n$^{++}$ junction.
As already observed from I-V and C-V measurements, the addition of Carbon, with low concentration, in the gain layer
volume slightly reduces the active doping concentration, and the reduction is more evident in sensors with Gallium. The addition
of a high concentration of Carbon results in a strong reduction of the gain layer, bringing the gain of those sensors to be $\sim$ 1.
Also, it is visible the different diffusion of Gallium with respect to Boron, with a broader and deeper peak, resulting in a lower peak
concentration. It will be shown in Sec.~\ref{sec:gain}, that despite the lower peak concentration, the electric field and thus the gain of
sensors with Gallium is higher than the one with Boron, as the Gallium implant is deeper and the high electric field region is
broader.~Referring to the Boron only case, it is interesting to notice how the different diffusion temperatures
impact on the width and the height of the doping peak.~Keeping the implantation dose constant (see Table~\ref{tab:split}),
Boron Low Diffusion is the configuration resulting in highest gain, for a given reverse bias.

\section{Laboratory Measurements on FBK UFSD2 Production}

Laboratory measurements are performed through a Transient Current Technique (TCT) setup, inducing a signal into the sensors using an infrared picosecond
laser, with a wavelength of 1060 nm. The laser can be focused with a lens and the minimum diameter of the laser spot is 10 $\mu$m.
The sensor output signal is amplified by a factor 100 using a Cividec Broadband Amplifier, a low-noise current amplifier
with an analog bandwidth of 2 GHz, and readout by a Lecroy Oscilloscope (740Zi) with a sampling of 40 GSample/s and a bandwidth of 4 GHz.

\subsection{Gain Measurement}
\label{sec:gain}

In the 50 $\mu$m UFSD FBK production, standard silicon diodes with no gain implant (PiN) are paired to each 1~$\times$~1~mm$^2$ UFSD single pad.
The PiN diode is used to measure the gain of the UFSD, by shooting both the PiN and the UFSD with the infrared laser calibrated
at the same intensity and using the relation:
\begin{equation}
  G = UFSD~Signal~Area~/~PiN~Signal~Area~.
\label{eq:G}
\end{equation}

\begin{figure}[t!]
\centering
\includegraphics[width=1\linewidth]{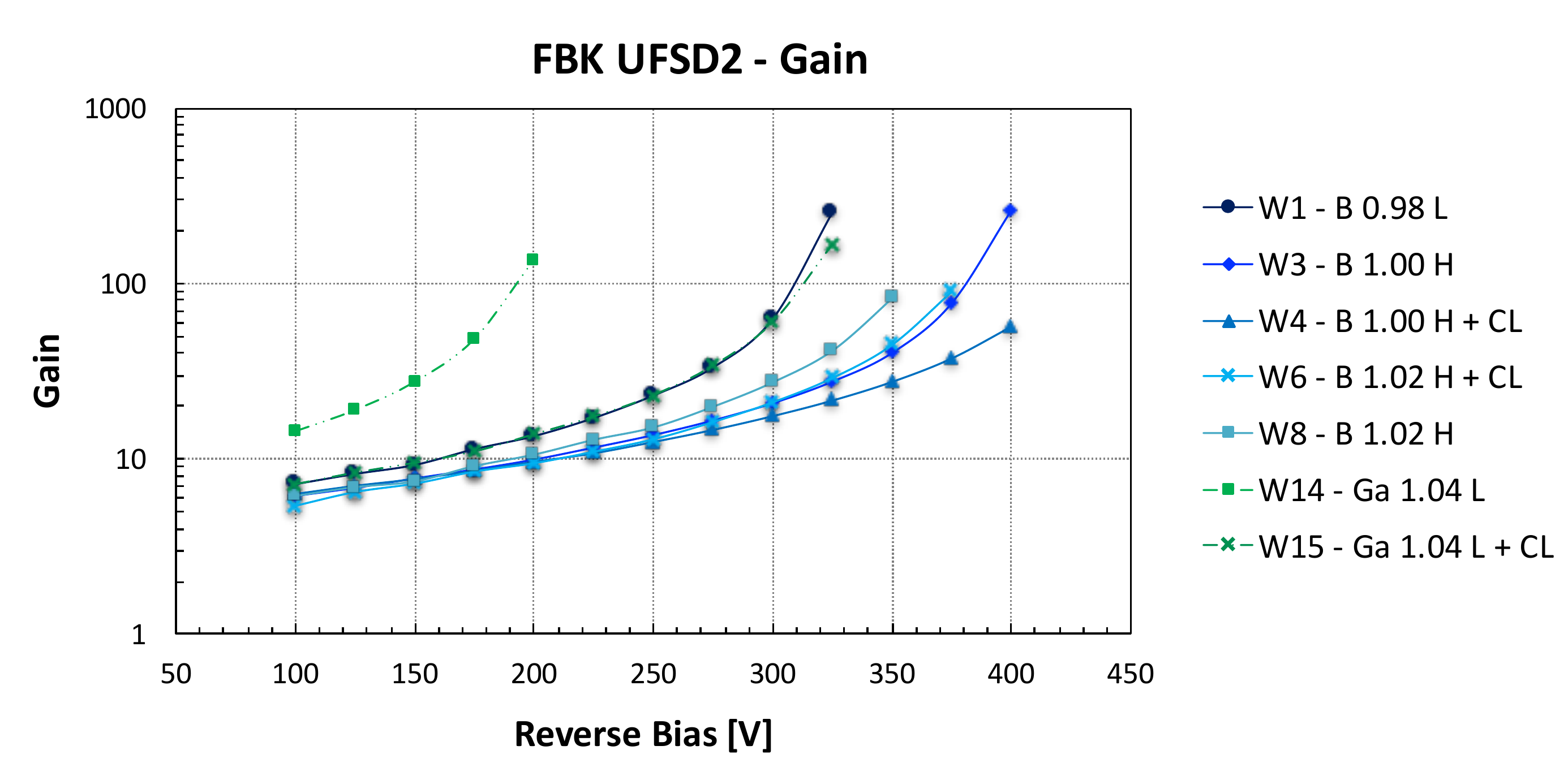}
\caption{Gain measurement as a function of the applied reverse bias for UFSD with different gain
  layer strategies: Boron Low Diffusion, Boron, Boron + Low Carbon, Gallium and Gallium + Low Carbon.
  Lines are added to interpolate between the measured points.}
\label{fig:gain}
\end{figure}

Results of the gain measurement are shown in Fig.~\ref{fig:gain}. The exponential dependence of the gain from the electric field
(as mentioned in Sec.~\ref{sec:IV}), and thus on the applied reverse bias, is visible.
Two important observations emerge from the measurement:
\begin{itemize}
\item[i] Boron implant obtained with low temperature of diffusion has a higher gain with respect to the Boron with a high diffusion temperature,
  due to the different profile of the gain implant, as predicted by the simulation~\cite{MM-IEEE17} and shown in Fig.~\ref{fig:DPb};
\item[ii] The presence of Carbon in the gain implant volume reduces the gain, and the reduction is more evident in
  sensors with a Gallium implant, as already anticipated in Sec.~\ref{sec:DC}.
\end{itemize}

\subsection{Time Resolution Measurement}

The time resolution of a detector, $\sigma_t$ , results from the sum of different contributions~\cite{UFSD1}, (i) the Jitter, $\sigma_{Jitter}$,
(ii) the Landau Time Walk, $\sigma_{Time Walk}$, the Landau noise, $\sigma_{Landau~Noise}$ and (iv) the signal distortion $\sigma_{Distortion}$:
\begin{equation}
  \sigma_t^2 = \sigma_{Jitter}^2 + \sigma_{Time Walk}^2 + \sigma_{Landau~Noise}^2 + \sigma_{Distortion}^2 ~.
\end{equation}
For the results presented here, the following simplification can be applied:
\begin{itemize}
\item[i] The effect of time walk can be compensated using a Constant Fraction Discriminator (CFD) analysis;
\item[ii] In silicon sensors the shape of the signal can be calculated using Ramo's theorem, which states that the signal induced by a charge
  carrier is proportional to the drift velocity of the charge, $v_{drift}$, and to the weighting field inside the sensor volume, $E_w$,
  according to $i \propto q \cdot v_{drift} \cdot E_w$, where $q$ is the charge generated in the detector volume.~The
  signal distortion can be neglected if the drift velocities of electrons and holes are saturated and if the weighting field is very uniform.
\end{itemize}
Assuming these simplifications, the predominant terms of the time resolutions are Jitter and Landau noise.

\subsubsection{Jitter}
\label{sec:jitter}

\begin{figure}[t!]
\centering
\includegraphics[width=0.8\linewidth]{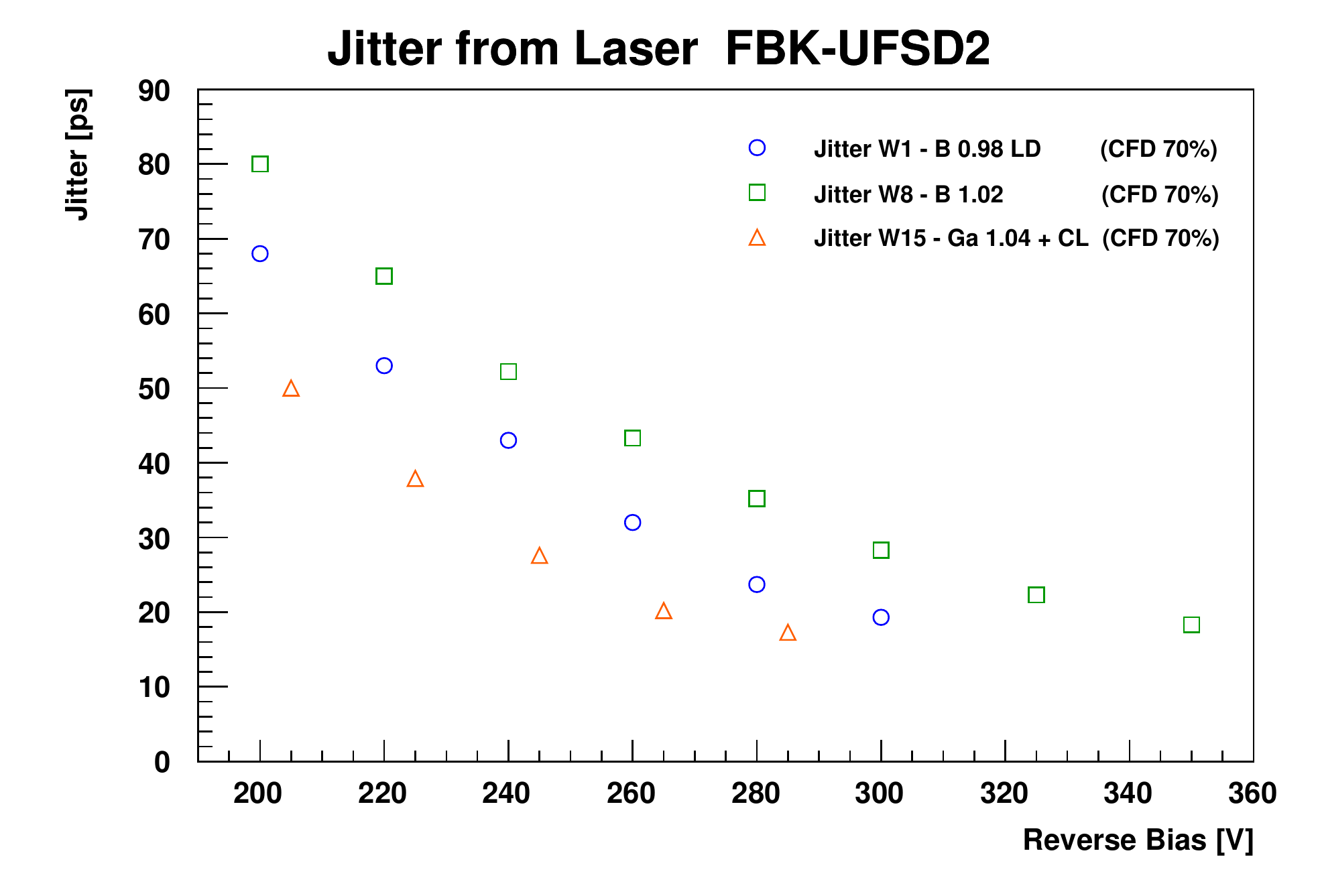}
\caption{Jitter as a function of the applied reverse bias measured on sensors with three different gain layer configurations:
  Boron Low Diffusion (blue dots), Boron (green squares) and Gallium + Low Carbon (orange triangles).}
\label{fig:jitter}
\end{figure}

The jitter term represents the time uncertainty caused by the presence of noise, and it is influenced by the signal steepness.~In particular,
it is directly proportional to the noise, $N$, and inversely proportional to the slope of the signal, $dV/dt$: $\sigma_{Jitter} = N / (dV/dt)$.

Figure~\ref{fig:jitter} shows the measurement of the jitter as a function of the bias voltage, obtained using the TCT setup and calibrating the laser intensity
to replicate the most probable value of the charge generated by a MIP passing through the sensor.~The jitter decreases increasing the reverse bias,
and therefore the gain of the sensor, reaching the value of 20 ps at the highest voltages\footnote{The setup used for timing measurement introduces a
  contribution to the timing resolution, as the distance between the sensors mounted on the PCB and the amplifiers is more than 1 cm, and this long path of the signal
  to the amplification add a parasitic capacitance that worsen the timing performance of the system. It could be possible to gain $\sim$ 5 ps using a more performing
  setup, with no parasitic capacitance.}.~The measurement of the jitter has been performed on sensors with three different configurations of the gain
layer: (i) Boron Low Diffusion, (ii) Boron, and (iii)
Gallium + Low Carbon.~These results show that different doping strategies of the gain layer do not affect the timing performance of the sensors,
and setting the proper reverse bias according to the different gain layer implants brings to the same ultimate jitter of $\sim$ 15 ps.

\subsubsection{Landau Noise}

\begin{figure}[b!]
\centering
\includegraphics[width=0.8\linewidth]{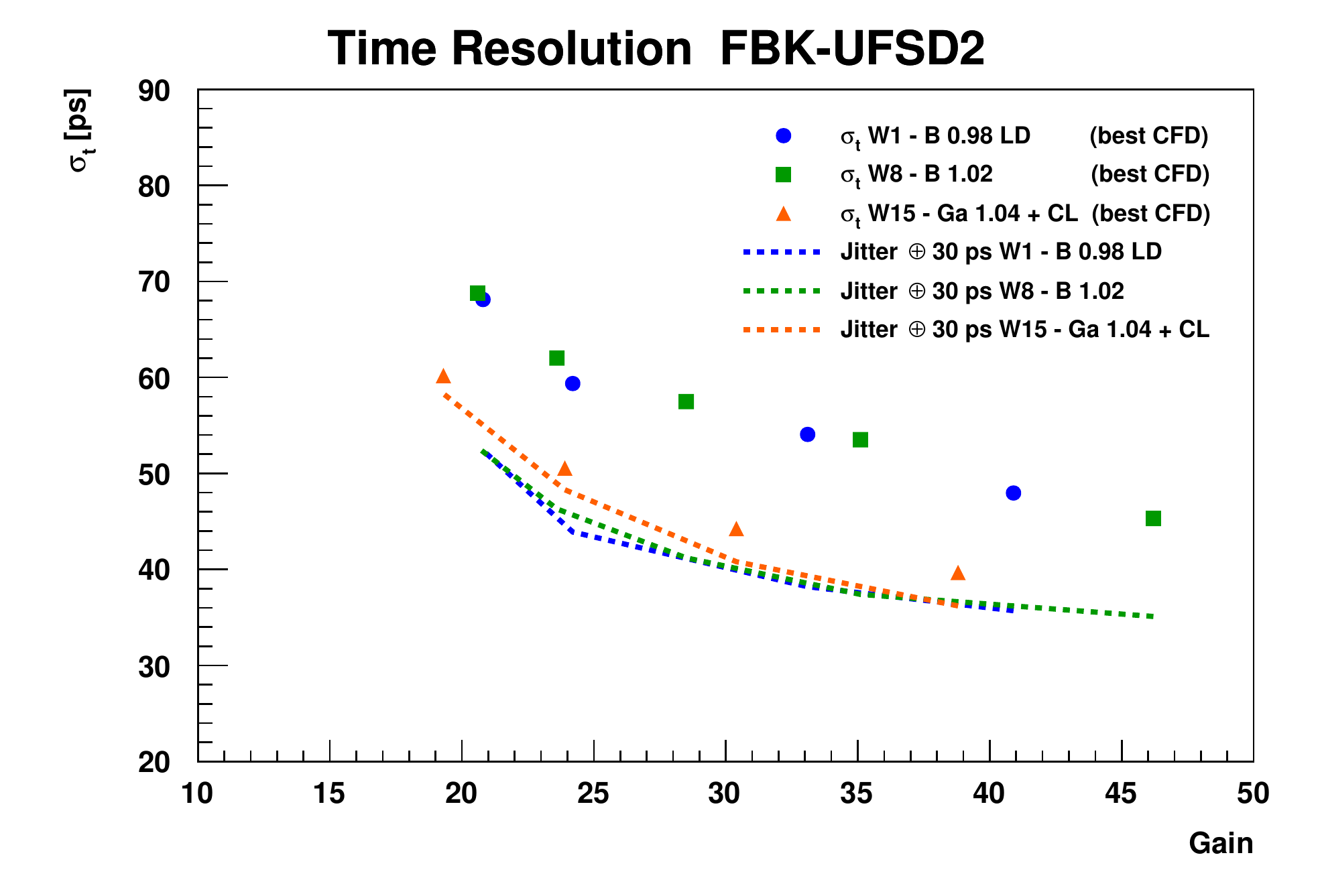}
\caption{Time resolution as a function of the gain measured on sensors with three different gain layer configuration: Boron Low Diffusion (blue dots),
  Boron (green squares), and Gallium + Low Carbon (orange triangles). Jitter contributions plus the predicted Landau noise of $\sim$ 30 ps
  are also shown as dotted lines.}
\label{fig:tres}
\end{figure}

Ionising particles crossing a silicon sensor create different energy depositions on an event-by-event basis, determining amplitude and
shape variations on the current signals, that follow the Landau statistics.~Landau noise contribution decreases with sensor thickness
and reaches $\sim$ 30 ps for 50 $\mu$m thick detectors.

Time resolution from MIP crossing the UFSD has been measured in a beam test at CERN SPS~\cite{TB-SPS}, using a beam of 180 GeV/c pions. The
trigger used for the data acquisition was a previously calibrated 50~$\mu$m thick UFSD~\cite{UFSD3}, with a time resolution of 35 ps.
The time resolution measured making use of a CFD analysis is shown in Fig.~\ref{fig:tres} as a function of the gain,
for the same sensors presented in Sect.~\ref{sec:jitter},
with gain layer configurations based on (i) Boron Low Diffusion, (ii) Boron, and (iii) Gallium + Low Carbon. Superimposed to the results, also the 
measured jitter terms (Fig.~\ref{fig:jitter}) plus the Landau noise contributions of $\sim$ 30 ps, as predicted for 50 $\mu$m thick sensors~\cite{UFSD1},
are shown.

Only the sensor with Gallium + Low Carbon configuration reaches the expected time resolution of $\sim$ 40 ps at the highest gain ($\sim$ 40). The two
sensors with Boron gain layer have a further contribution that worsen the overall time resolution by $\sim$ 10 ps.~The reason might be related to the
hole effect that will be discussed in Sec.~\ref{sec:holeff}.~More information on this contribution will come from beam tests with a tracking system.

From the obtained results it is possible to assert that the addition of Carbon in the gain layer volume does not affect the timing performance
of the UFSD.

\subsection{The Hole Effect}
\label{sec:holeff}

\begin{figure}[b!]
\centering
\includegraphics[width=0.7\linewidth]{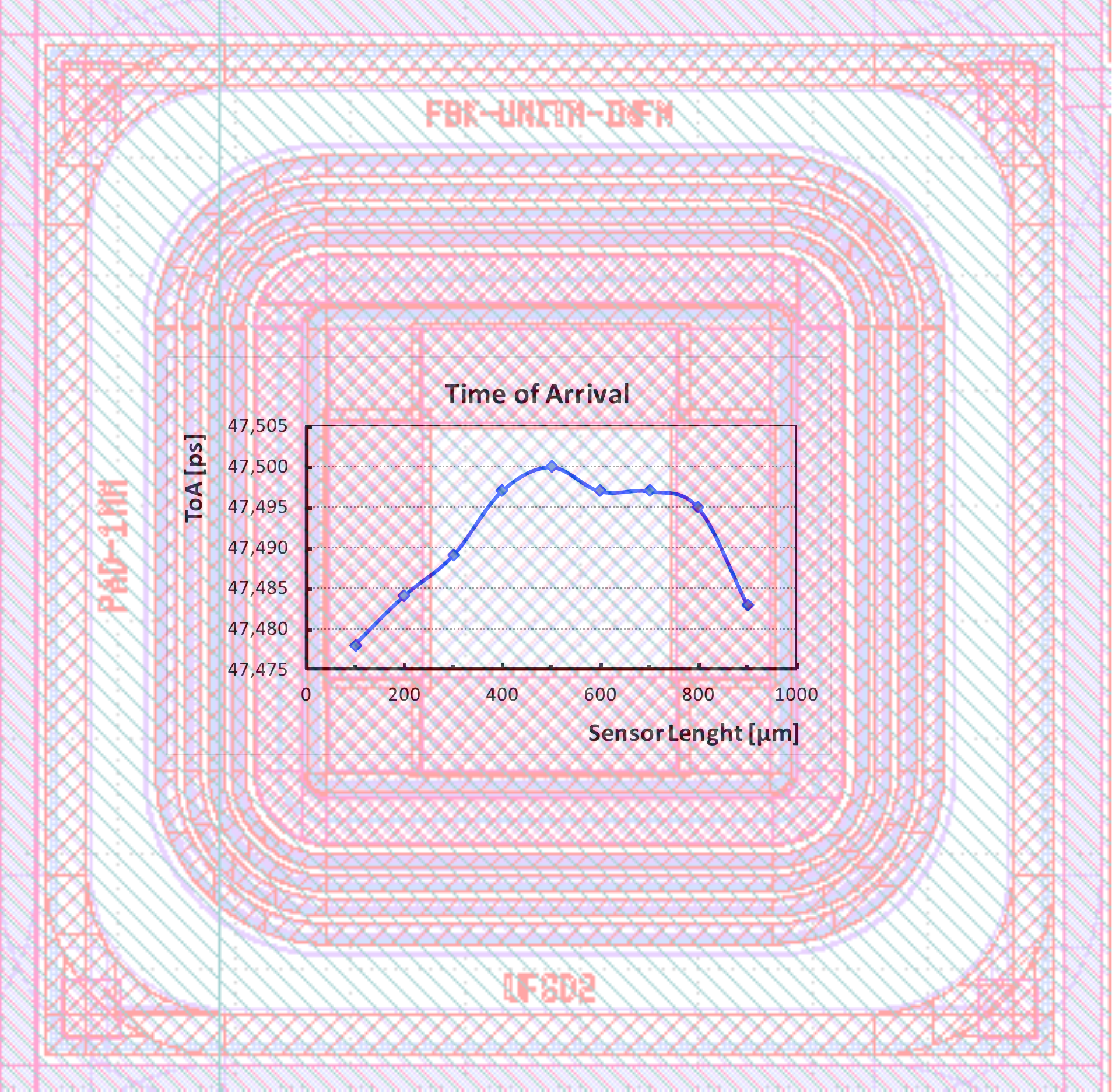}
\caption{Time of arrival of the signal along the length of the sensor, using the edge-TCT technique. The CAD image of the 1 $\times$ 1 mm$^2$
  pad is superimposed. According to the figure, the laser is shot from the top margin of the sensor.}
\label{fig:ToA}
\end{figure}

Most of the sensor structures designed for the FBK UFSD2 production, have an opening on the metallisation
that contact the n$^{++}$ ohmic implant, the so-called optical window (hole), to allow for laser tests on the sensors.

It has already been observed on devices manufactured by other foundries~\cite{NC-RD50}, that the amplitude and the time of arrival (ToA)
of the signals from the two different regions, the one covered by the metal and the one with the opening, differ.~More in detail,
the signals generated on the volume below the opening region have a smaller amplitude ($\sim$ 10 mV less)
and arrive tens of picoseconds later than the signal generated on the sensor volume covered by the metal.

To investigate if also the FBK UFSD2 production suffers the same amplitude and ToA differences between the metal-non metal regions,
an edge-TCT scan has been performed on a 1 $\times$ 1 mm$^2$ sensor, confirming the results observed on sensors from other
producers.~Results on time of arrival are reported on Fig.~\ref{fig:ToA}: the
signals from the region with the opening on the metallisation arrive
$\sim$ 20 ps later than the signals generated below the metallisation\footnote{The time of arrival is defined as the difference between the time given
  from the trigger signal of the laser and the time of the signal generated inside the sensor, both taken at 50\% of their amplitude.}.~The
effect still need to be understood and it is not reproduced by device simulation, but for the UFSD usage in experiments sensors fully
covered by metallisation are foreseen.

A small ToA asymmetry with respect to the middle of the detector might be due to a little ($\sim$ 1 $\mu$m over the full sensor lenght)
misalignment between the sensor plane and the laser system.

\subsection{Intra-Pad Inactive Region}

\begin{figure}[ht!]
\centering
\includegraphics[width=0.7\linewidth]{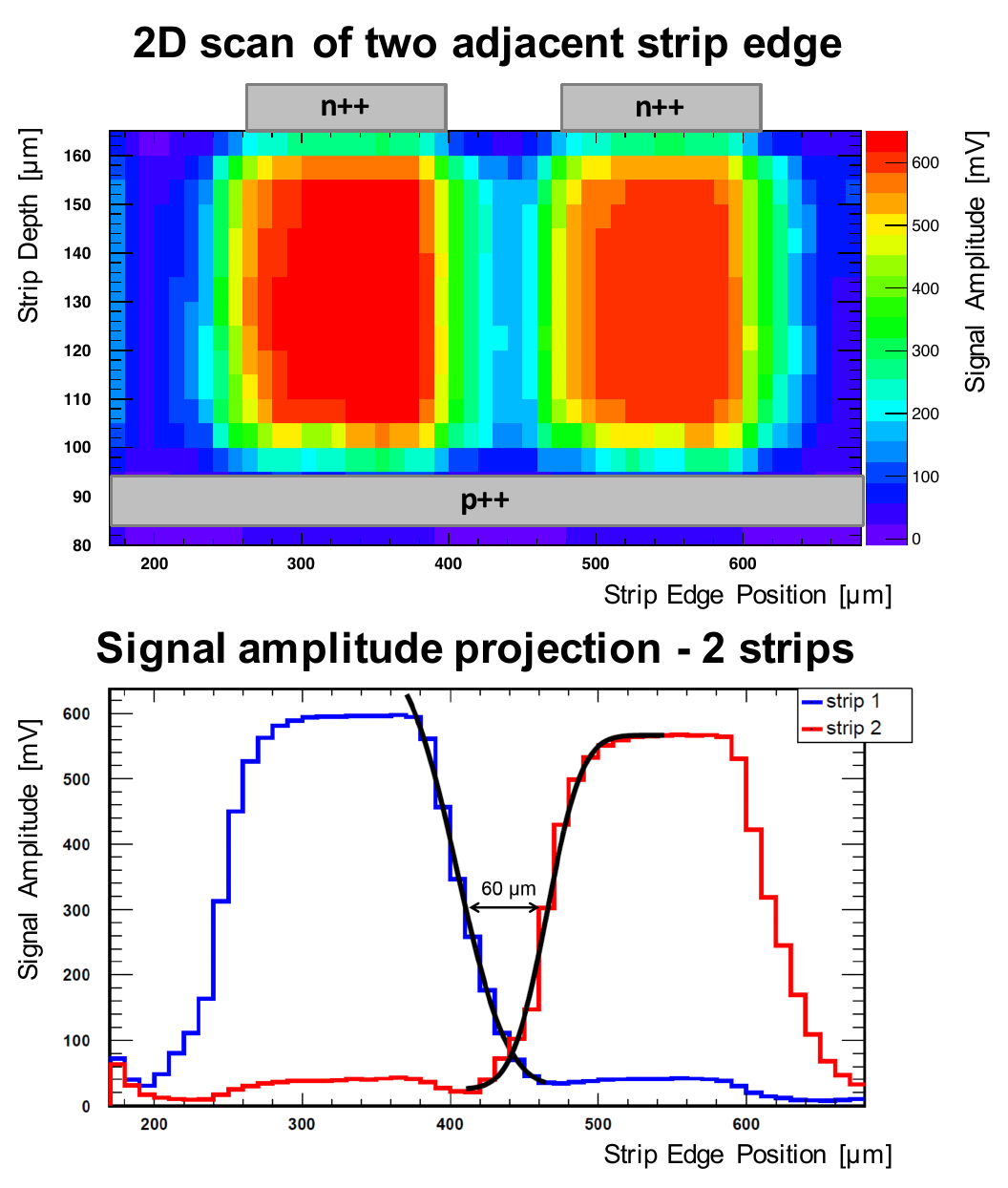}
\caption{Edge-TCT scan on a multi-strip sensor. Top: 2D scan of the sensor edge, strip position on the x axis and strip depth on the y axis.
  Bottom: 1D projection of the signal amplitude along the strip position, the amplitude as read from two different strips is shown.}
\label{fig:IntraPad}
\end{figure}

Pixelated and multi-strip UFSD structures need to have a good gain termination at the edge of each pixel.~An important
parameter for such structures is the size of the inactive region between two adjacent active pads.~The measurement of
the inactive region magnitude in FBK UFSD2 production has been performed on strip sensors using the edge-TCT
technique.~The strips under test have a width of 150 $\mu$m and a pitch of 200 $\mu$m.

Figure~\ref{fig:IntraPad} shows the results of the edge-TCT:~from the 2D plot (top) is visible the separation between two adjacent
strips, showing that the two strips are well isolated, while the 1D plot (bottom) of the signal amplitude readout from two different
channels projected along the p++ sensor electrode (see top plot for reference) shows
that the full width of the inactive region at half height of the maximum amplitude is about 60 $\mu$m,~in
agreement with the expectations from the sensor layout.

\section{Radiation Tolerance of FBK UFSD2 Production}

FBK UFSD2 sensors have been irradiated with neutrons at the JSI TRIGA reactor in Ljubljana~\cite{JSI-Lj} and with 24 GeV/c protons at the IRRAD CERN
irradiation facility~\cite{CERN-Irr} to test their radiation tolerance and identify the most radiation resistant strategy for the gain layer
(for more details, see~\cite{UFSD2irr}).~Sensors with different gain layer configurations have been irradiated:~(i) Boron Low Diffusion,
(ii) Boron, (iii) Boron + Low Carbon, (iv) Gallium, and (v) Gallium + Low Carbon.~Investigated wafers and fluences are summarised in Table~\ref{tab:irr}.

\begin{table}[h!]
\begin{center}
\begin{tabular}{ | c | c | c | c | c | } 
  \hline
  Wafer $\#$ & Dopant & Dose & n fluence [10$^{15}$ n$_{eq}$/cm$^2$] & p fluence [10$^{15}$ p/cm$^2$] \\
  \hline
  1  & B LD       & 0.98 & 0.2, 0.4, 0.8, 1.5, 3.0, 6.0 & 0.6, 3.0  \\
  6  & B + C Low  & 1.02 & 0.2, 0.4, 0.8, 1.5, 3.0, 6.0 & 0.6, 3.0  \\
  8  & B          & 1.02 & 0.2, 0.4, 0.8, 1.5, 3.0, 6.0 & 0.6, 1.0, 3.0, 6.0 \\
  14 & Ga         & 1.04 & 0.2, 0.4, 0.8, 1.5, 3.0, 6.0 & 0.6, 3.0  \\
  15 & Ga + C Low & 1.04 & 0.2, 0.4, 0.8, 1.5, 3.0, 6.0 & 0.6, 3.0  \\
  \hline
\end{tabular}
\caption{Summary of wafers and fluences used in the irradiation campaign.}
\label{tab:irr}
\end{center}
\end{table}

UFSD devices under irradiation suffer for removal of acceptor atoms in the gain layer volume~\cite{GK-Irr1}.
The initial acceptor removal, $N_A$, is exponentially dependent on the irradiation fluence, $\phi$, according to:
\begin{equation}
  N_A(\phi) \propto N_A(0) \cdot e^{- c \frac{ \phi }{ \phi_0 } } ~,
\label{eq:removal}
\end{equation}
where $N_A(0)$ is the initial doping concentration, $\phi_0 = $ 1 particle/cm$^2$ is a normalisation factor,
and $c$ is the acceptor removal coefficient.

\begin{figure}[b!]
\centering
\includegraphics[width=0.8\linewidth]{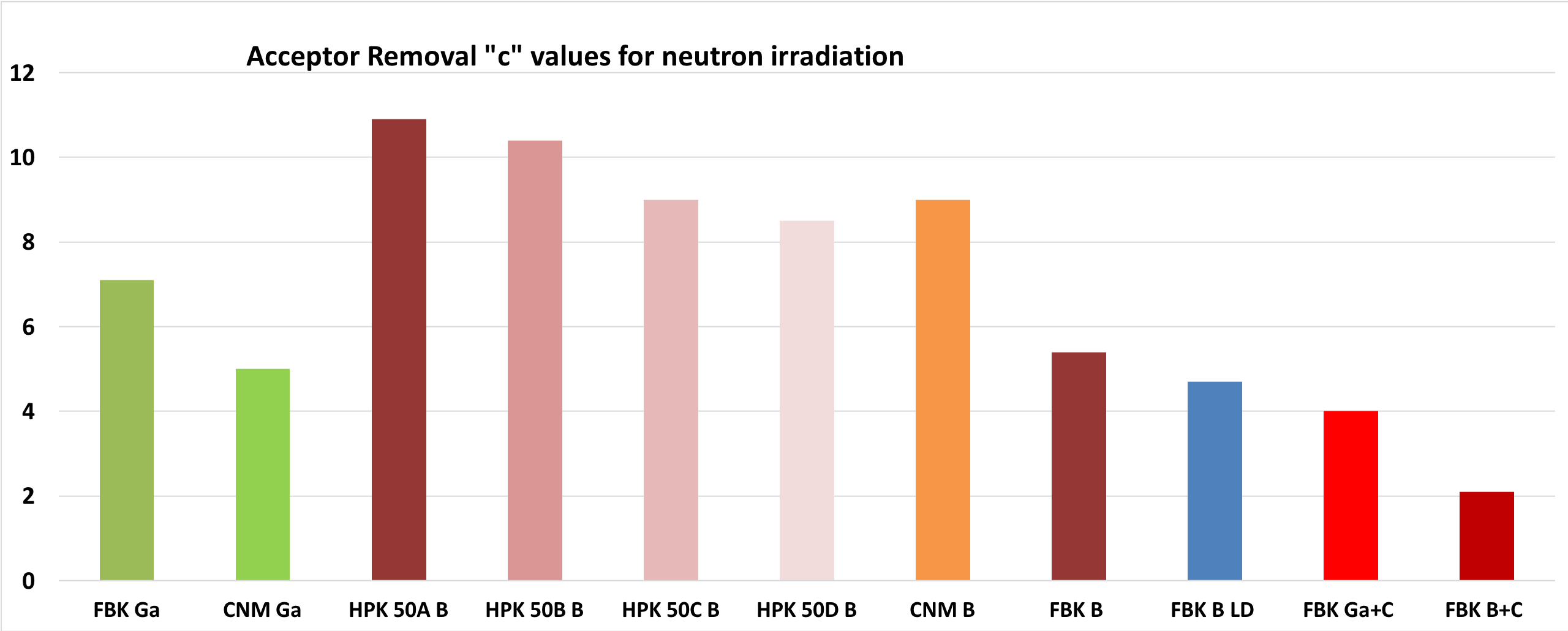}
\caption{Initial acceptor removal coefficient $c$ from neutron irradiation of UFSD
  manufactured by 3 different foundries (CNM, FBK, and HPK) with different gain layer doping compositions.~Effective~$c$ values
  are multiplied by 10$^{-16}$.}
\label{fig:cnCoef}
\end{figure}

To estimate the reduction of the gain layer dopant dose, the removal coefficient $c$ from Eq.~\ref{eq:removal} has been studied.
The survival active acceptor concentration has been estimated using the C-V measurement: from the voltage at which the gain layer
is depleted (drop of the capacitance, as shown in Fig~\ref{fig:CVall}), it is possible to estimate the fraction of gain layer dopant
still active in the sensors, as a function of the received fluence.
By an exponential fit to the data points, the $c$ coefficient has been extracted. 

Figure~\ref{fig:cnCoef} shows the values of the initial acceptor removal coefficient, $c$, for sensors irradiated with neutrons.
Together with the 5 different configurations of the gain layer for the sensors produced by FBK,
namely Boron Low Diffusion, Boron, Boron + Low Carbon, Gallium, and Gallium + Low Carbon,
the plot reports also measurements for sensors manufactured by CNM, with Boron and Gallium,
and by HPK, with 4 different doses of Boron implant~\cite{GK-Irr2}. The value of $c$ is higher for sensors with Gallium, meaning that
those sensors are more prone to suffer from initial acceptor removal under irradiation, while the lowest values of $c$ are reached
for sensors with Low Carbon concentration underneath the gain implant. In particular, the configuration Boron + Low Carbon
is the most radiation tolerant, more than a factor of two with respect to the case without Carbon.~Among sensors without Carbon,
the gain implant with Boron diffused at low temperature is the most radiation hard.

\begin{figure}[b!]
\centering
\includegraphics[width=0.9\linewidth]{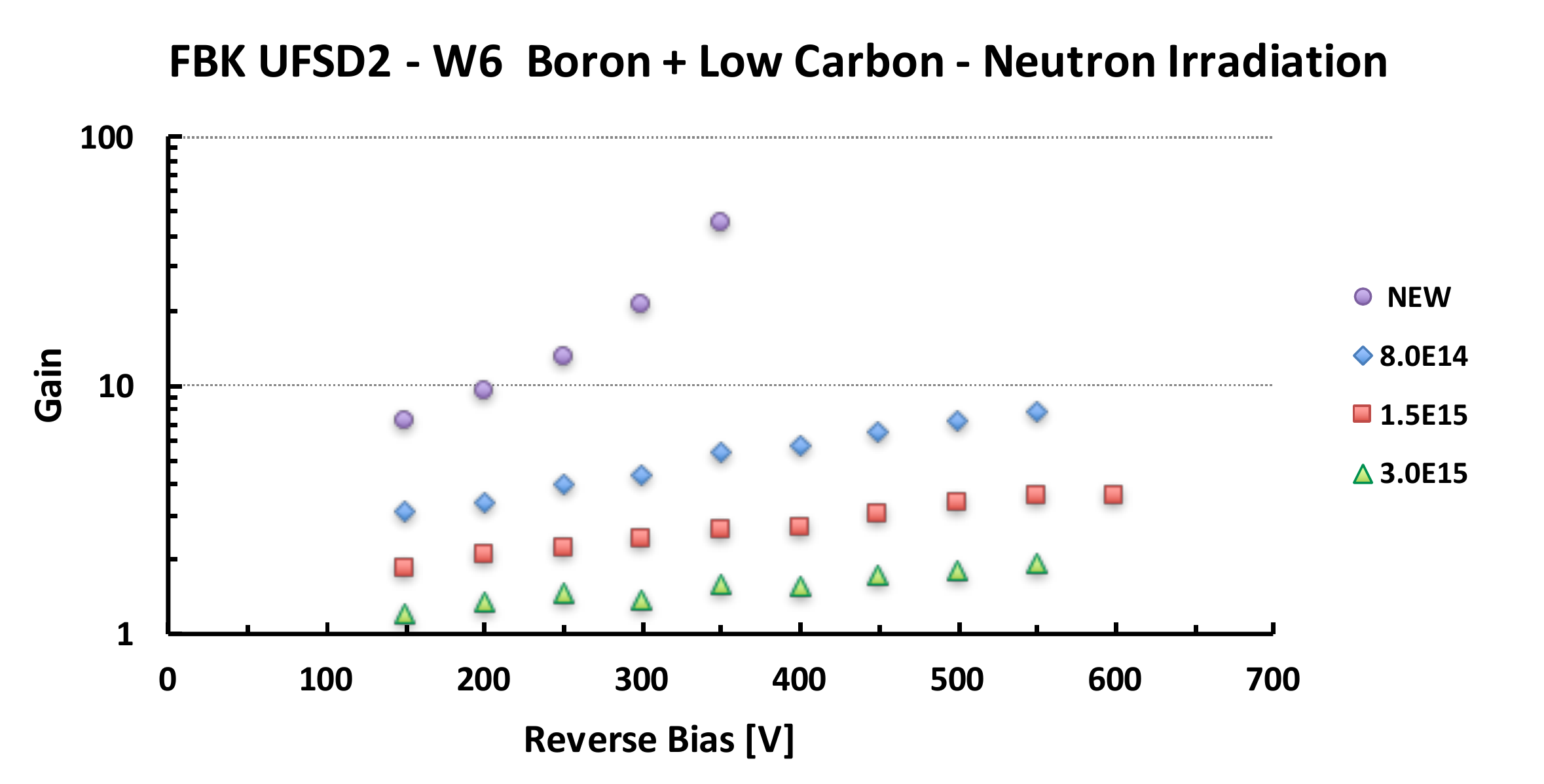}
\caption{Gain due to the gain layer implant for sensors with Boron + Low Carbon after reactor neutron irradiation,
for a not-irradiated sensor and for three irradiation fluences: 8.0E14, 1.5E15, and 3.0E15 n$_{eq}$/cm$^2$.}
\label{fig:gainW6i}
\end{figure}

Besides the study of the acceptor removal coefficient, the gain due to the survived gain layer implant has been studied, for all the configurations
tested under irradiation. During the irradiation campaign both PiN diodes and UFSD have been irradiated at the same fluences, and the
signal multiplication due to the gain layer contribution only is defined as:
\begin{equation}
  G = Irradiated~UFSD~Signal~Area~/~Irradiated~PiN~Signal~Area~.
\label{eq:Girr}
\end{equation}
Figure~\ref{fig:gainW6i} shows the gain due to the gain implant contribution only for sensors with Boron plus Low Carbon concentration
in the implant volume, for the sensor before irradiation and after reactor neutron irradiation at three different fluences, namely 8.0E14, 1.5E15, and 3.0E15
1 MeV equivalent neutrons per cm$^2$ (n$_{eq}$/cm$^2$). 
It is possible to notice that even after 3.0E15 n$_{eq}$/cm$^2$, there is still a contribution to the signal multiplication due to the
gain layer implant. Also, the irradiation extends the range of reverse bias at which it is possible to operate the detector, making accessible
a region where the electric field in the bulk region is high enough to obtain charge multiplication along the whole sensor volume
(see e.g.~\cite{HPKirr} for more details).

It has been observed that 24 GeV/c protons produce the same amount of initial acceptor removal as a function of the fluence (particles/cm$^2$)
as neutrons~\cite{UFSD2irr}.

\section{Conclusion}

The first 50 $\mu$m thick UFSD production at FBK has been delivered.
Several options for the gain layer implant have been pursued, in particular (i) Boron Low Diffusion, (ii) Boron, (iii) Boron plus Carbon,
(iv) Gallium, and Gallium plus Carbon have been used to obtain an moderate internal charge multiplication.

Extensive tests and electrical characterisations on wafer show that the production is of very high quality.
Further tests in laboratory prove that the internal gain, the timing performance and the extension of the inactive region between pads
are as expected.
Irradiation tests with neutrons and protons, assert that implanting Carbon atoms underneath the gain layer implant volume doubles the radiation
resistance of the UFSD detectors, while for the configurations without Carbon, the Boron implant diffused at low temperature is the one
responding better under irradiation.

Given the success of this first 50 $\mu$m thick UFSD production at FBK, the future plans are to further improve the radiation
hardness of UFSD and to prove the capability to produce large area UFSD, to make FBK a good candidate for the
sensor production for timing detectors foreseen for the ATLAS and CMS upgrade.

\section*{Aknowledgments}

We thank our collaborators within RD50, ATLAS and CMS who participated in the development of UFSD. Our special thanks to the technical
staff at INFN Torino and FBK Trento.~Part of this work has been financed by the European Union Horizon 2020 Research and Innovation
funding program, under Grant Agreement no.~654168 (AIDA-2020) and Grant Agreement no.~669529 (ERC UFSD669529), and by the
Italian Ministero degli Affari Esteri and INFN Gruppo V.

\section*{References}

\bibliography{vs_bibfile}

\end{document}